# Backward wave optical parametric oscillation in a waveguide

PATRICK MUTTER[1,2], FREDRIK LAURELL[1], VALDAS PASISKEVICIUS[1] AND ANDRIUS ZUKAUSKAS[1]

[1]*Department of Applied Physics, Royal Institute of Technology (KTH), Roslagstullsbacken 21, Stockholm 10691, Sweden*
[2]*Svenska Laserfabriken AB, Ruddammsvägen 49, Stockholm 11419, Sweden*
*\*pmutter@kth.se*

**Abstract:** A backward wave optical parametric oscillator (BWOPO) waveguide in periodically poled Rb-doped KTP is presented. The waveguide exhibits low loss (0.16 dB/cm) and has an oscillation threshold, almost 20 times lower than the corresponding bulk device. The backward wave has a narrow linewidth of 21 GHz at 1514.6 nm while the forward wave at 1688.7 nm has a spectrum replicating the pump. The unique spectral features of the BWOPO will unlock novel opportunities in low-power nonlinear integrated optics. A conversion efficiency of 8.4% was obtained limited by the emergence of backward stimulated polariton scattering.

## 1. Introduction

Integrated nonlinear optics enables next-level advancements for a wide range of applications, including signal processing, wavelength multiplexing, laser sources, quantum information processing, quantum communications, and photonic quantum computing because of the possibility to create miniaturized compact devices, potential to integrate with existing photonic platforms, and improved efficiency of the nonlinear processes due to high intensities provided by the intrinsic confinement of the electromagnetic field and long interaction lengths. [1,2]. Quasi-phase matching (QPM) attracts special interest in the realm of integration because of the possibility to tailor nonlinear optical processes to a further degree, offering flexibility for precise tuning of such properties as nonlinearity, output frequency, spectral content, etc. Advantages of QPM have been extensively exploited lately, particularly in the quantum optics community where ultrahigh efficiency devices were demonstrated [3,4]. The QPM technique relies on a periodic modulation of the nonlinearity with a period $\Lambda$ to compensate the process' phase mismatch, $\Delta k = m\frac{2\pi}{\Lambda}$, where $m$ is the order parameter [5]. Early work focused on QPM waveguides in $LiNbO_3$ [6,7], but van der Poel *et al.* showed that segmented $KTiOPO_4$ (KTP) waveguides was a viable alternative combining a better mode overlap between the interacting modes and a higher power handling capability [8,9]. For the segmented waveguides the ion-exchange induced the domain reversal and automatically formed the QPM structure. An extensive review of the topic can be found in ref [10].

    The backward wave optical parametric oscillator (BWOPO) is a down-conversion device where one of the generated waves propagates counter-directional to the pump, while the other travels co-directionally [11]. In contrast to a conventional optical parametric oscillator (OPO), which requires external feedback through an optical cavity, the BWOPO automatically establishes distributed feedback due to the counter-propagating nature of the nonlinear interaction. The lack of physical cavity endows the BWOPO with robustness, ease of alignment and high efficiency in a single-pass configuration, all within a compact footprint. Conversion efficiencies exceeding 70% and outputs at the mJ level have recently been achieved [12]. Furthermore, a BWOPO offers drastically different spectral properties compared to its co-propagating counterpart. The forward wave inherits the spectrum and phase of the pump, while the backward-generated wave is inherently spectrally narrow [13–15]. These unique spectral features are gaining attention in the quantum optics community as they allow to generate high-purity and narrowband heralded single photons [16–18]. Despite the lack of the filtering effect

of a cavity, high level of squeezing, comparable to conventional OPOs, should be achievable when operated in the vicinity of the oscillation threshold [19]. The above-mentioned properties makes the BWOPO an interesting candidate for integration in quantum optical systems.

The antiparallel propagation of signal or idler photon with respect to the pump introduces a huge phase-mismatch $\Delta k$ represented as,

$$\Delta k = k_p - k_f + k_b = m\frac{2\pi}{\Lambda}, \qquad (1)$$

where $k$ is the wavevector and the subscripts $p, f$ and $b$ denote pump, forward and backward waves, respectively. This large phase-mismatch makes QPM the sole, viable method for satisfying the phase-matching condition. For first order ($m = 1$) interactions QPM periods below 1 µm are required in classical ferroelectrics like $LiNbO_3$, $LiTaO_3$ or KTP. However, the fabrication of such fine-pitched QPM structures presents a formidable task, and is the primary reason why it took such a long time to achieve the first experimental demonstration [13]. Recent advancements [20,21] have shown that the limitations in fabrication can be surmounted through coercive field engineering in rubidium-doped KTP (RKTP), allowing consistent fabrication of bulk sub-µm domain gratings with periodicities presently down to 317 nm [22]. This method utilizes ion exchange to establish alternating regions with low and high coercive fields, enabling polarization reversal through electric field poling using planar electrodes. Nevertheless, the integration of a waveguide in a BWOPO remains an unexplored avenue albeit the promise of benefits due to the field confinement over long interaction lengths. Studies involving counter-propagating nonlinear interactions in waveguides have, until now, been constrained to high-order ($m=3$) QPM with periodicities of 1.7 µm [23] and 1.3 µm [24]. The scenario of shorter QPM periodicities being fabricated first in bulk compared to waveguide devices may appear counterintuitive considering that QPM gratings within waveguides should be more feasible due to significantly reduced aspect ratio requirements. The reason is that the most efficient quasi-phase-matched waveguides in KTP were fabricated by: first implementing the QPM grating trough electric field poling and subsequently inscribing the waveguide via ion exchange [10]. However, the delicate nature of finely pitched ferroelectric domains renders them vulnerable to the elevated temperature that is necessary for ion exchange [25]. The reduced nonlinearity associated with higher-order quasi-phase-matching made it impossible to reach BWOPO threshold in a waveguide so far.

In this work we present sub-micrometer ($\Lambda$ = 409 nm) periodically poled RKTP waveguides and use them to demonstrate the first-ever waveguide BWOPO. The waveguides were fabricated in a two-step process: first a segmented ion-exchanged periodic structure was formed on a polar surface of RKTP crystal, which served both as the waveguides and a coercive field grating. Afterward, the QPM grating was inscribed via electric field poling, eliminating the need for any subsequent heat treatment.

## 2. Results

The waveguide PPRKTP sample with 55 waveguides was placed in the BWOPO setup illustrated in Fig. 11. The waveguides were 22 mm long with a QPM grating length of 20 mm. A Ti:sapphire regenerative amplifier was used as the pump laser, providing linearly chirped pulses with a duration of 215 ps and a chirp rate of 19.4 mrad/ps$^2$, at a repetition rate of 1 kHz. The pulses had a center wavelength of 798.9 nm and a spectral bandwidth of 1.0 nm. A waveplate-polarizer arrangement was used to control the launched pump energy and ensured polarization aligned along the polar axis of the crystal to exploit the large $d_{33}$ coefficient. Light was coupled into the waveguide using an objective lens (Melles Griot, 06 GLC 003) with a focal length of 8.48 mm and 0.28 numerical aperture (N.A.). A second objective lens (Melles Griot, 06 GLC 001) with a focal length of 1.21 mm and N.A. 0.61 was used to collect the light after the waveguide. Dichroic mirrors positioned before and after the waveguide were used to separate the pump from the down-converted light. Transmission measurements below the oscillation threshold reveal 92% waveguide transmission over a length of 22 mm for the pump, which corresponds to a low loss of 0.16 dB/cm.

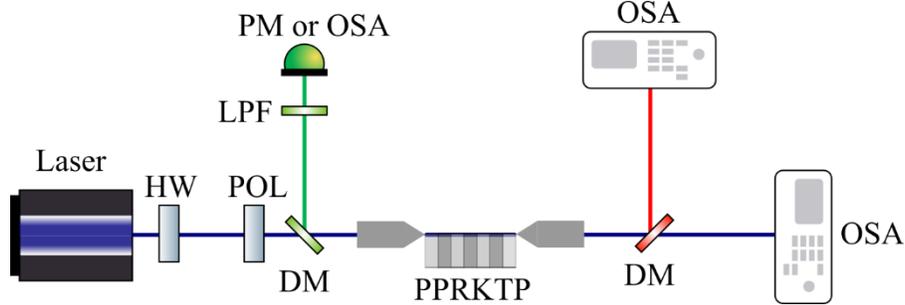

Fig. 1: Experimental setup for optical evaluation of the BWOPO waveguide. HW, half-wave plate; POL, polarizer; DM, dichroic mirror; LPF, long-pass filter; PM, power meter; OSA, optical spectrum analyzer; PPRKTP, periodically poled RKTP sample.

The BWOPO threshold was reached in 50 out of the 55 waveguides present on the sample. We opted to conduct a more comprehensive investigation of a waveguide with a width of 9.8 µm. For this waveguide, the BWOPO threshold was reached for a pump energy of 325 nJ, representing a remarkable 19-fold reduction when compared to recent bulk BWOPO experiments conducted with the same laser used as a pump source [26]. Fig. 2 displays the spectra of the forward and backward waves recorded at pump energies corresponding to twice the oscillation threshold. For comparison, spectra generated in the bulk of another sample with the same QPM grating period are shown as well. Due to waveguide dispersion, the central wavelength of the backward wave shifted from 1511.7 nm (bulk) to 1514.6 nm (waveguide) while the forward wave shifted from 1692.8 nm (bulk) to 1688.7 nm (waveguide). The observed spectra exhibit the characteristic features of a BWOPO, with the pump phase being transferred to the forward wave, resulting in a broad spectrum (98 GHz), while the backward wave remaining spectrally narrow (21 GHz).

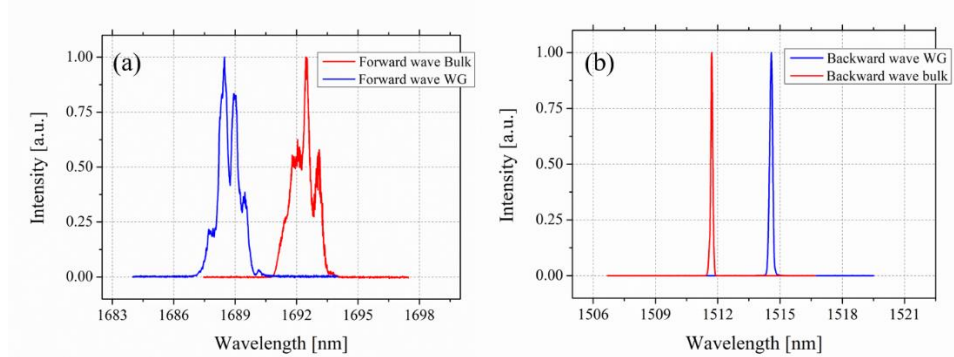

Fig. 2: Forward (a) and backward (b) wave spectra at pump energies twice the oscillation threshold generated in waveguide and bulk, respectively.

Fig. 3 presents the output energy of the parametric waves and the conversion efficiency as a function of pump energy. Owing to difficulties measuring low energies at 1688.7 nm, only the backward wave energy was directly measured, while the forward wave's energy was determined using the Manley-Rowe relations. At a pump energy of 714 nJ a peak conversion efficiency of 8.4% was observed. At larger pump energies the BWOPO efficiency declined due to the emergence of backward stimulated polariton scattering (BSPS). In KTP, BSPS is known to be a very efficient nonlinear process that can compete with the quasi-phase matched nonlinear processes [27]. In fact, the first BWOPO demonstration suffered from limited conversion efficiency for the same reason [13]. BSPS is not supressed by the QPM grating since the slow polariton wave does not experience the χ(2) modulation before absorption takes over. In a well-poled bulk PPKTP sample, the parametric nonlinear interaction has a lower threshold and higher gain than BSPS, therefore, the pump energy is channeled into the parametric interaction, and the BSPS process does not reach its threshold. However, with an imperfect QPM grating the effective nonlinearity for the three-wave-mixing process is reduced, and the BSPS process can compete with the down-conversion process and partially deplete the pump energy. Additionally, for a waveguide the modal overlap area, $A_{ovl}$, is an important parameter, as the efficiency, $\eta$, scales as $\frac{d_{eff}^2}{A_{ovl}}$. A large overlap area will hence put higher demands on the QPM structure quality.

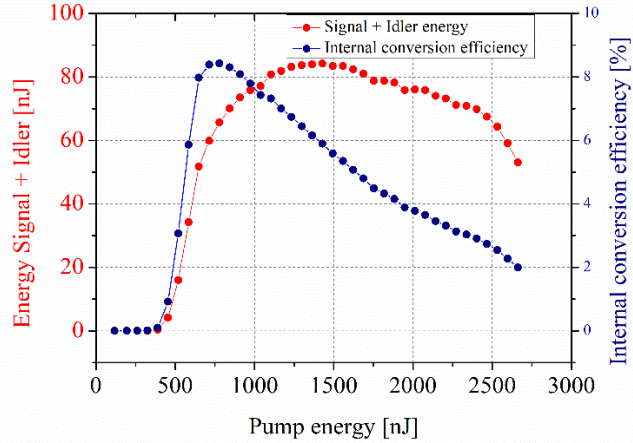

Fig. 3: Internal conversion efficiency and energy of the parametric waves as a function of the pump energy.

To investigate this further we conducted simulations of the waveguide mode profile using the wave optics module in COMSOL. To ensure the accuracy of these simulations, it was crucial to have precise refractive index data. For that reason, we subjected a single-domain reference RKTP sample to planar ion exchange with the same recipe as for the segmented waveguides. Energy dispersive X-ray spectroscopy was employed to determine the exchange profile. As expected, the Rb$^+$-ion concentration follows a complementary error function [28], with an erfc(z)-depth of 13 µm. The planar waveguides were measured via prism coupling [29] with a He-Ne laser at 633 nm and the surface refractive index change was calculated to be 0.0064 using the inverse Wentzel-Kramer-Brillouin method [30]. Given that the BWOPO waveguides are segmented (duty cycle ~50%), the surface refractive index change for the waveguides was modelled as half of the measured value of the planar waveguide. Fig. 4 displays the simulated and normalized intensity distribution of the TM$_{00}$ modes of the pump-, backward- and forward waves for the RKTP waveguide. Due to the asymmetric refractive index profile the intensity maximum pushed deeper in the RKTP waveguide for longer wavelengths. With such suboptimal field overlap the nonlinear interaction is reduced. We calculate the overlap area as;

$$A_{ovl} = \frac{\iint E_p^2(x,z)\,dx\,dz \iint E_b^2(x,z)\,dx\,dz \iint E_f^2(x,z)\,dx\,dz}{\left[\iint E_p(x,z)E_b(x,z)E_f(x,z)\,dx\,dz\right]^2}, \qquad (2)$$

where $E_p$, $E_b$ and $E_f$ are the electric field distribution of the pump, backward and forward modes, respectively. Notably, an increased field overlap would result in a smaller effective area and a lower BWOPO threshold. Utilizing the electric field distributions obtained from the COMSOL simulation and applying them to equation (2) results in an effective area of 296 µm². Substituting this value into the threshold equation derived in [31], we ascertain the effective nonlinear coefficient $d_{eff} = 3.5 \frac{\text{pm}}{\text{V}}$ (ideally 10.8 $\frac{\text{pm}}{\text{V}}$ in KTP [32]). It confirms a non-optimal QPM structure (see, Fig 5).

To enhance the efficiency of the BWOPO in waveguides, it is essential to lower the oscillation threshold to avoid emergence of BSPS. Our results highlight two key avenues for potential improvements. Firstly, there is a need to further enhance the quality of the QPM structure. Despite excellent control over the domain duty cycle, double domain reversal via ion-exchange in the exchanged regions and electric field poling in non-exchanged regions effectively nullifies the first 3-5 µm of the QPM structure. Improvements in this context could involve identifying an ion exchange procedure that creates a large enough refractive index increase to guarantee waveguiding and simultaneously, either avoids or deepens the polarization reversal. The latter would eliminate the need for electric field poling. The second avenue for improvement involves increasing the modal overlap, a goal that could be accomplished by using buried waveguides. One viable approach might entail employing a secondary ion exchange process in a $KNO_3$ melt to reverse the effects of the initial exchange at the top layer, akin to the reverse proton exchange technique utilized in $LiNbO_3$ [33] and $LiTaO_3$ [34]. However, it is unclear how this would impact the coercive field grating.

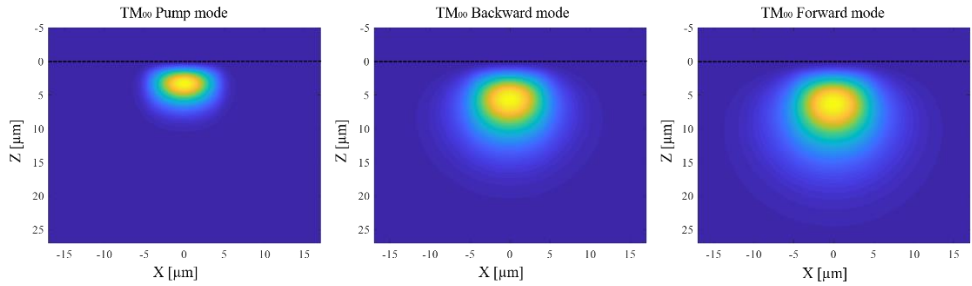

Fig. 4: Intensity mode profiles of the $TM_{00}$ modes of the pump-, backward- and forward waves. The dashed black line denotes RKTP surface positioned at z=0, with the positive z-direction indicating the bulk and the negative z-direction representing air.

In summary, we have presented a reliable waveguide and QPM fabrication technique for RKTP featuring a QPM period well below 1 µm. It was used to demonstrate the first-ever waveguide BWOPO. The technique should be scalable to even shorter QPM periods, enabling down-conversion with both signal and idler counter-propagating to the pump - a nonlinear interaction that has not yet been demonstrated. The waveguides exhibited low loss (0.16 dB/cm), however, the conversion efficiency was limited to 8.4% due to the onset of BSPS. We have also outlined strategies to improve conversion efficiency. Already now the unique spectral features of counter-propagating down-conversion should allow for efficient generation of high-purity and narrowband single photons in a ready for integration package, offering exciting prospects in quantum optics.

## 3. Method

To form the QPM waveguides a 24 mm long RKTP sample was patterned on the c⁻-surface with a 409 nm period photoresist grating using UV-interference lithography. Next, an ion exchange stop layer was introduced within the openings of the photoresist by exposing the patterned surface to oxygen plasma etching [35]. We then removed the remaining photoresist and patterned the same surface with a waveguide pattern using conventional photolithography to form channels along the x-direction with widths from 6 µm to 22 µm. Subsequently, we subjected the RKTP surface once more to oxygen plasma etching. Thereafter, we performed the ion exchange by immersing the crystal into a molten salt bath containing 20 mol% KNO$_3$, 73% RbNO$_3$ and 7 mol% Ba(NO$_3$)$_2$, at a temperature of 330 °C for 4 hours. This ion exchange recipe is well studied for electric field poling via coercive field engineering [21]. As a result, segmented ion exchange channels were formed that served both as waveguides and coercive field gratings. In the final step we periodically poled the segmented structure by contacting the two polar surfaces with planar liquid electrodes and applying 1.25 ms long pulses of 8 kV.

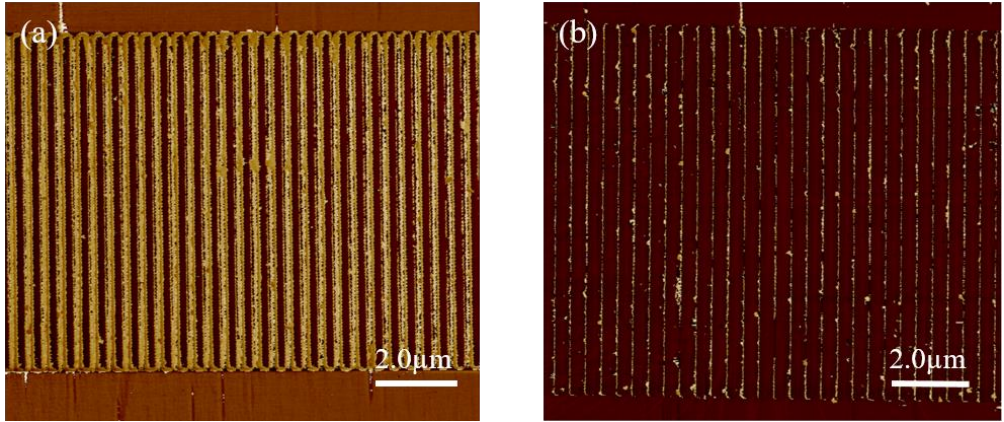

Fig. 5. PFM images of the polar surface of two different waveguide regions. In (a) only the electric field poling inverted the polarization. In (b) the polarization was inverted via electric field poling and ion exchange.

Fig. 5 shows piezo force microscopy (PFM) images of two representative examples of the ferroelectric domain structure for a 10 µm wide waveguide. In Fig. 5.5 (a) the polarization is inverted in the non-exchanged areas only, resulting in a nearly ideal QPM structure with 50% duty cycle. However, in most of the waveguides, the ferroelectric domain structure was erased, as illustrated in Fig. 5.(b). We attribute it to two independent mechanisms: electric field poling in the non-exchanged regions, and spontaneous domain reversal in the ion exchanged regions during the ion-exchange process. The latter was exploited in the original method for making QPM waveguides in KTP [8]. In our case, only extremely thin domains, approximately 40 nm thick, at the interface between exchanged and non-exchanged regions remained unaltered, as can be seen in Fig. 5.5 (b). To obtain further insight into the ferroelectric domain structure, the y-surface of a second sample was polished and scanned with the PFM, see Fig. 6. It was processed with the same recipe as described above, but patterned with a period of 580 nm. As one can see at the top of the sample, domains inverted via ion exchange are cone-shaped with depths ranging from 3 to 6 µm. The domains inverted via electric field poling maintain their width for approximately 25 µm before merging. The ion exchange prevents domain merging at shallower depth, aligning well with findings in ref [21]. The strong confinement of the polarization switching to the non-exchanged areas renders this technique well-suited for waveguide fabrication. Although the applied voltage was too high for periodic poling of the RKTP bulk, the waveguide area exhibits significant resilience towards higher electric fields.

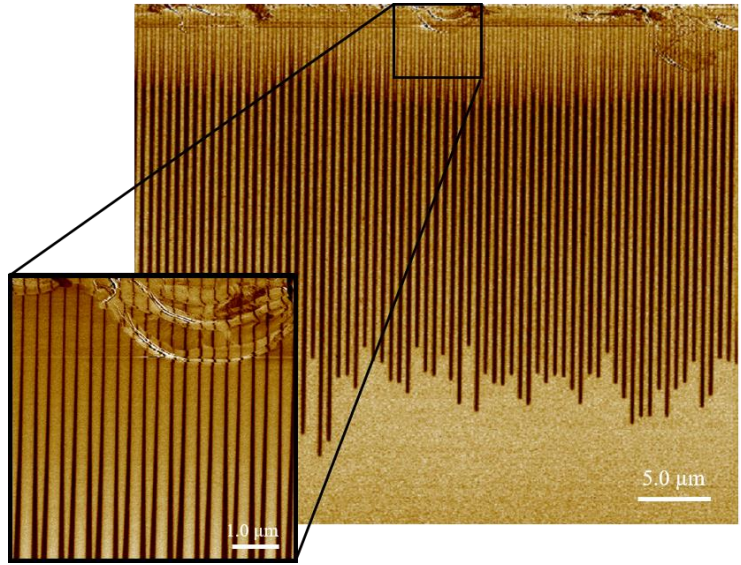

Fig. 6: PFM image of the ferroelectric domain structure along the polar axis (y-face). The domains inverted via electric field poling propagate approximately 25-30 µm into the bulk before they merged. The depth of the domains inverted via ion exchange exhibit cone-shaped structures with depths ranging from 3 to 6 µm.